\documentstyle[12pt]{article}

\begin{document}

\begin{center}{{\bf
TRIALITY IN QCD AT ZERO AND FINITE TEMPERATURE: A NEW DIRECTION
               }\\
\vglue 1.0cm
{M.~Faber}\\
\vglue 0.2cm
\baselineskip=14pt
{\it Institut f\"ur Kernphysik, Technische Universit\"at Wien,}\\
\baselineskip=14pt
{\it Wiedner Hauptstr. 8-10, A-1040 Vienna, Austria}\\
\vglue 0.4cm
{O.~Borisenko, G.~Zinovjev}\\
\vglue 0.2cm
\baselineskip=14pt
{\it Institute for Theoretical Physics, Ukrainian Academy of Science,
Kiev 143}\\
\vglue 1.4cm
{ABSTRACT}}
\end{center}
\vglue 0.6cm

Discrete symmetries in grand canonical ensembles and in ensembles
canonical with respect to triality are investigated. We speculate about
the general phase structure of finite temperature gauge
theories with discrete $Z(N)$ symmetry. Low and high temperature phases
turn out to be different in both ensembles even for infinite systems.
It is argued that gauge theories with matter fields in the fundamental
representation should be treated in ensembles canonical with
respect to triality if one wants to avoid unphysical predictions.
Further, we discuss as a physical consequence of such a treatment the
impossibility of the existence of metastable phases in the quark-gluon plasma.

\newpage

The recent interest in studying the $Z(3)$ structure of hot
gauge theories \cite{preprint}-\cite{boor} is caused by the strong will to
obtain more insight into the phase structure of strongly interacting matter
for physical models and for speculations. In the present article we follow the
obvious
tendency of the above papers to avoid a polemic with each other, to supplement
the conventional approach \cite{Mac_Lerran} and to develop our own point of
view \cite{preprint,L94}. The guideline of this exploration is to draw
the arguments supporting the conclusion of \cite{preprint,Polonyi} which
come from the impressing recent activity on the role of local and global
discrete symmetries in gauge theories \cite{ds1,ds2,ds3,ds4}.
Usually, these symmetries appear as a commuting centre of
an underlying gauge group. An important conclusion of the above papers
is an appearance of a new Higgs phase where the corresponding
centre charges are not screened while a gauge symmetry is broken
up to a discrete subgroup. Moreover, unlike in the confined phase
these charges could, in principle, be detected at long distances
via the Aharonov-Bohm effect. In order to make our arguments
more clear we have first to review some aspects of the present status of MC
analysis of finite temperature QCD.

\section{$Z(3)$ symmetry in the grand canonical ensemble description
(GCE). Euclidean formulation}

In lattice QCD the gluon fields appear in the form of $SU(3)$ matrices
$U_{x,\mu}$ which are defined on links $(x,\mu)$ of a four-dimensional
euclidean lattice. In finite temperature lattice QCD the correlation
function $<L(\vec{r}_1) \cdots L^*(\vec{r}_N)>$ of several Polyakov loops

\begin{equation} \label{Polyakov_loop}
L(\vec{r}) = \frac{1}{3} \; {\rm Tr} \prod_{t=1}^{N_t} U_{0}(\vec{r},t)
\end{equation}

\noindent can be connected \cite{Mac_Lerran} with the free energy $F$ of $N$
infinitely heavy quarks $q$ or antiquarks $\bar{q}$ at the corresponding
positions $\vec{r}_1,\cdots , \vec{r}_N$ at temperature $T$ relative to the
free energy of the vacuum

\begin{equation} \label{free_energy}
F(q(\vec{r}_1),\cdots ,\bar{q}(\vec{r}_N)) = -T \, {\rm ln}<L(\vec{r}_1)
\cdots L^*(\vec{r}_N)>.
\end{equation}

\noindent The thermodynamical average $< \cdots >$ is computed using the
partition function

\begin{equation} \label{path_integral}
Z = \int{} {\cal D} [U] \; e^{-S[U]}
\end{equation}

\noindent as a ''sum'' over all gauge field configurations $U$. In pure
gluonic QCD for most problems the appropriate choice for $S$ is the Wilson
action

\parbox{11cm} {\begin{eqnarray*}
S_G[U] & = & \beta \sum_{x, \mu < \nu} \left(1 - \frac{1}{3} {\rm Re} \;
{\rm Tr} \; U_{x, \mu \nu} \right), \quad \beta = \frac {6}{g^2},\\
U_{x, \mu \nu} & = & U_{x, \mu} U_{x+\hat{\mu}, \nu} U^{\dag}_{x+\hat{\nu},
\mu} U^{\dag}_{x, \nu} \, ,
\end{eqnarray*}} \hfill
\parbox{1cm}{\begin{eqnarray} \end{eqnarray}}

\noindent where the plaquettes $U_{x, \mu \nu}$ are built of four links in
the $\mu \nu$-plane of a four dimensional euclidean lattice.

The action $S_G[U]$ is $Z(3)$ symmetric. This means that multiplication of
all links in direction $\mu=0$ in the three dimensional $x,y,z$-torus with
fixed $t$, e.g. $t=0$ by a $Z(3)$ element leaves the action invariant. Such
a global $Z(3)$ transformation can be regarded as an aperiodic
transformation of the centre of the gauge group. A single
Polyakov loop $L(\vec{r})$ transforms under $Z(3)$ nontrivially
and therefore the distribution of $L(\vec{r})$ values can be used
as an indicator for spontaneous breaking of $Z(3)$ symmetry.

Let us now discuss the results of lattice calculations and their common
interpretation. Monte-Carlo calculations in pure gauge lattice QCD show usually
a
characteristic behaviour of the spatial average $L$ of $L(\vec{r})$ during
a Monte-Carlo simulation. In the confinement regime (low $\beta$) $L$
scatters symmetrically around zero in the complex plane. In the deconfined
phase there appear three $Z(3)$ symmetric maxima of the Polyakov loop
distribution in $0^{\circ}$- and $\pm 120 ^{\circ}$-directions. The
appearance of the three peak structure of the $L$-distribution in the
deconfinement regime close to the phase transition is considered as a
demonstration that spontaneous symmetry breaking on a finite lattice can
never happen exactly. As the number of tunnelling events between the maxima
decreases with increasing $\beta$ one commonly expects $L$ to be frozen in
the thermodynamical limit in one of the $Z(3)$ directions and thus to get
spontaneous $Z(3)$ symmetry breaking. Therefore, one may obtain ${\rm
arg}<L> = 1, \pm \frac{2 \pi}{3}$ in the thermodynamical limit.
With dynamical fermions the action $S$ contains a fermionic
contribution $S_F$

\parbox{11cm} {\begin{eqnarray*}
&& S = S_G + S_F, \\
&& S_F = \frac{n_F}{4} \sum_{x,x^{\prime}} \bar{\Psi}_x M_{x,x^{\prime}}
\Psi_{x^{\prime}}, \ M_{x,x^{\prime}} = D_{x,x^{\prime}} + m
\delta_{x,x^{\prime}}
\end{eqnarray*}} \hfill
\parbox{1cm}{\begin{eqnarray} \end{eqnarray}}

\noindent which breaks $Z(3)$ symmetry explicitly. In the Kogut-Susskind
formulation \\
\cite{staggered_fermions} the fermionic matrix M reads

\parbox{11cm} {\begin{eqnarray*}
&& M_{x,x^{\prime}} = \frac{1}{2} \sum_{\mu} \left( \Gamma_{x, \mu} U_{x,
\mu} \delta_{x^{\prime}, x+\mu} - \Gamma_{x^{\prime}, \mu}
U^{\dag}_{x^{\prime}, \mu} \delta_{x^{\prime}, x-\mu} \right) + m
\delta_{x,x^{\prime}}, \\
&& \Gamma_{x, \mu} = (-1)^{x_1+x_2+ \cdots +x_{\mu-1}},
\end{eqnarray*}} \hfill
\parbox{1cm}{\begin{eqnarray} \end{eqnarray}}

\noindent with one-component Grassmann variables $\Psi_x$ and
$\bar{\Psi}_x$. The integration over the Grassmann variables in the path
integral

\begin{equation}
Z = \int{} {\cal D} [U,\bar{\Psi},\Psi] \; e^{-S}
\end{equation}

\noindent can be performed analytically and leads to the fermionic
determinant ${\rm det} M$

\begin{equation} \label{fermionic_path_integral}
Z = \int{} {\cal D} [U] \; e^{-S_G[U]} {\rm det} M.
\end{equation}

It is easy to see that the fermionic action violates $Z(3)$ symmetry
\cite{Banks}. In the usual MC-iterations one finds that in the low
temperature phase the expectation value $<L>$ of a Polyakov loop $L$ is a
small positive number and in the high temperature phase the three peak
structure of the distribution becomes asymmetric. The maximum at $0^\circ$
is favoured and the two lower maxima at angles of around $\pm120^\circ$ are
symmetric. This fact is usually interpreted as a consequence of the
violation of $Z(3)$ symmetry in QCD.

Thus, the lore is:

In the low temperature phase $Z(3)$ symmetry signals the confinement of colour
charges and in the high temperature phase spontaneous breakdown of $Z(3)$
symmetry means colour charges screened by gluon fields and the Polyakov line
is an appropriate order parameter. In QCD with dynamical quarks,
$Z(3)$ symmetry is explicitly broken and Polyakov loop ceases
to be the appropriate order parameter. It should be stressed
that these results reflect the grand canonical ensemble treatment of full QCD
with respect to triality.

These results are grounds for nonperturbative MC analysis of QCD providing
a theoretical basis of quark-gluon plasma phenomenology. However, this
treatment is not
complete and absolutely satisfactory. There exist several reasons
to believe in such a conclusion. This approach does
not solve the long standing problem of an order parameter for full QCD.
Moreover, it fails in tempting to treat $Z(3)$ metastable phases
as physical states in Minkowski space
due to running into thermodynamical inconsistency. We shall argue that this
picture is conceptually not perfect as well, since it does not substantiate
unified mechanism of quark confinement at zero and finite temperature.
Indeed, confinement at zero temperature implies only zero
triality states and unbroken $Z(3)$ symmetry  which triality is
connected with.  Meanwhile, the
corresponding symmetry is explicitly broken in full QCD at any low
temperature.  At least, it does not answer the important question what
occurs if the triality charge is not screened and develops - according
to the appropriate Gauss law - a long-range observable which
could be, in principle, experimentally measurable.

We are going to demonstrate below that there are no such problems
in the canonical ensemble description of finite temperature QCD.

\section{$Z(3)$ symmetry in the canonical ensemble description
(CE). Euclidean formulation}

We develop a view \cite{preprint,Polonyi,polonyi1} different from
that mentioned above which seems more consistent at least in many
general aspects. Let us begin mentioning the Euclidean formulation of
the theory at zero temperature. The Lagrangian of QCD is symmetric
under transformations of the gauge colour group $SU(3)_{loc}$. There
is a global subgroup of the local symmetry which can be decomposed as
$SU(3)_{gl} = SU(3)/Z(3)_{gl} \otimes Z(3)_{gl}$. Gauge fields are
identically invariant under $Z(3)_{gl}$ transformations but fermionic
fields transform as
\begin{equation} \label{fermiontr}
\Psi (x) \rightarrow  e^{i \frac{2 \pi}{3}k} \Psi (x)
= e^{i \frac{2 \pi}{\sqrt{3}}\lambda^{8} k} \Psi (x)
\end{equation}
\noindent
leaving the QCD Lagrangian invariant.
This symmetry is related to triality
$t_q$ which is defined as
\begin{equation}
t_{q} = (N_{q} - N_{\bar{q}})mod~3
\label{triality}
\end{equation}
\noindent
or as expectation value of the operator
\begin{equation}
t_{q} = < \sum_{x}  \bar{\Psi} (x) \sqrt{3} \lambda^{8}  \Psi (x) >.
\label{troperator}
\end{equation}
\noindent
It had been proposed \cite{mack} that in QCD the gluon colour
charges and any zero triality charges are screened by a dynamical
Higgs mechanism (understood as a screening mechanism) leaving $Z(3)$
symmetry unbroken and a single quark (having nonzero triality)
unscreened. Then long-range chromoelectric forces between quarks will
persist according to the Gauss theorem for a discrete centre.

In the scheme described in the previous section (GCE) a single quark
may have a finite free energy and the states with open triality can
contribute to the partition function.
As a result the domains with different $Z(3)$
phases give different contributions to the partition function.
The zero phase contributes most.
We are sticking here to the opinion that this problem is an
artifact of the GCE description which leads to the explicit violation of
the $Z(3)$ aperiodic symmetry. We shall show below that these aperiodic
transformations at finite temperature can be regarded as acting
on links in time direction or equivalently on fermionic fields.
These finite temperature transformations of fermionic fields
must be treated in analogy to the zero temperature transformations
(\ref{fermiontr}). They are related to triality in the same manner
as (\ref{fermiontr}).
Since our premise is the statement that QCD with dynamical fermions
is able to describe only systems of zero triality as it is at zero
temperature and all quark phases should be indistinguishable both
in the partition function and in thermodynamical functions we would like
to construct a theory which is invariant under the corresponding
$Z(3)$ transformations.

Let us first describe how one can construct such a theory for pure
gluodynamics. In the absence of external perturbation (like
"magnetic"  field for spin systems), the average value of the Polyakov loop
will be equal to zero at any temperature even in the thermodynamical limit:
as $\int{} {\cal D} [U] \; e^{-S_G[U]}$ is $Z(3)$ invariant and the Polyakov
loop (\ref{Polyakov_loop}) is not, one has to get
\begin{equation} \label{vanishing_Polyakov}
<L> = 0
\end{equation}
in the confined and also in the deconfined phase. The important point is
there is no contradiction between this statement and the spontaneous
breaking of $Z(3)$ symmetry in the deconfined phase. To demonstrate that one
can use $<L(0) L^*({\vec r})>$ correlations or even more suitable the ($Z(3)$
symmetric)  $L(\vec{r})$ distributions expressed by the moments $|L({\vec
r})|^2, |L({\vec r})|^3, \cdots$, what is equivalent to the determination
of Polyakov loops  in $Z(3)$ invariant representations (octet, decuplet,
antidecuplet, etc) \cite{Multiplets}.

An obstacle for accepting $<L> = 0$ in the deconfined phase comes from
lattice calculations which give $<L> \ne 0$ at high $\beta$. This result
has to be compared with the usual method to demonstrate spontaneous
symmetry breaking by an external field. Including $- \lambda \sum_{\vec r}
Re\{zL(\vec r)\}, z \in \{ 1, e^{\pm 2 \pi i/3} \}$ in the QCD action,
performing the thermodynamic limit and then the limit to vanishing
$\lambda$ would result in $<zL> \ne 0$ and positive in the deconfined
phase. As we cannot influence a physical system by a field which we have
switched off we have to give an interpretation of this result: The external
field fixes the coordinate system in which the ''direction'' of spontaneous
symmetry breaking is measured. This direction is unchanged even after
switching off the external field. To get $<L> = 0$ we have to use
the ''unbiased'' path integral which averages over all directions of such
a coordinate system and gives $<L> = 0$ also in the deconfined phase.
The observation that for a $Z(3)$ invariant Lagrangian the observables of
nonvanishing triality have expectation value zero is rather trivial for the
pure gluonic case (see \cite{preprint,kiskis} for more details).
Nevertheless, the usage of the "unbiased" path integral (i.e., with
perturbation and averaging as described above) is more preferable
since we do not have violation of triality in such a treatment despite
deconfinement takes place. We will now turn to the more
interesting case of full QCD.

Periodic boundary conditions in imaginary time direction imply
that we are dealing with a theory defined in a space with topology
$R^{3} \otimes S^{1}$. To construct a theory with conserved triality
and indistinguishable phases of quark fields
we will follow a recently developed ideology
\cite{ds1,ds3} (see also \cite{ds2,ds4} and references therein).
Fields of gauge theories on space-time manifolds that are not simply
connected  need not to be strictly periodic on noncontractible
closed loops of this manifold. Instead, these fields need only be
periodic up to the action of an element of $Z(3)$. Just these boundary
conditions distinguish local $Z(3)$ symmetry from global $Z(3)$ symmetry
and allow the local symmetry to manifest itself through a nontrivial
Aharonov-Bohm effect.

Let us consider the Taylor expansion of $\exp (-S_F)$ in the path
integral. The first nontrivial term, coming from time-like links,
which will survive the integration over fermionic fields is proportional
to $\prod_{t}\bar{\Psi}_t \Psi_t U_0(t)$. This term describes propagation
of a single quark along the temperature direction. The triality
transformations
of the fermionic fields in  $\prod_{t}\bar{\Psi}_t \Psi_t U_0(t)$ should
give a nontrivial phase factor. It is obvious that one can achieve
this by considering  in addition to (\ref{fermiontr})
the following substitutions for the fermionic fields
\begin{equation} \label{fermiontr1}
\Psi (\vec{x}, \beta) \rightarrow - e^{i \frac{2 \pi}{3}k} \Psi (\vec{x}, 0).
\end{equation}
\noindent
If we now perform this substitution in the operator
$\prod_{t}\bar{\Psi}_t \Psi_t U_0(t)$ and then consider the action
of the transformation (\ref{fermiontr}) on this operator we will find
that the operator gets a nontrivial phase factor as it should be
for a state with a single quark. Thereby, this is the obvious extension
of triality transformations (\ref{fermiontr}) to the finite temperature
theory. The phase transformation (\ref{fermiontr}) determines the
triality of quark states at zero temperature. Hence, this transformation
together with (\ref{fermiontr1}) determines the triality of quark states
at finite temperature.

In the zero temperature theory the transformation (\ref{fermiontr})
acts trivially on all physical states. We would like to construct
a formulation of QCD where the corresponding finite temperature
transformations act trivially on physical states as well.
To get such a theory where only triality zero states contribute we
have to sum over the following boundary conditions in the time direction
\begin{equation}
\Psi (\vec{x}, \beta) = - e^{i \frac{2 \pi}{3}k} \Psi (\vec{x}, 0).
\label{boundcond}
\end{equation}
\noindent
This is equivalent to a summation over all phases in (\ref{fermiontr})
combined with (\ref{fermiontr1}) and to a projection onto zero triality
states.

We explain now that such a projection is nothing but the requirement
of local gauge invariance in the meaning of refs. \cite{ds2} and
\cite{ds3}. The boundary conditions
(\ref{boundcond}) allow to identify field configurations which are
related by the transformation (\ref{fermiontr1}). Such a relation can be
considered as the definition of a local discrete symmetry \cite{ds2}
and is known as the so-called orbifold construction.
It was shown \cite{ds2} that one must project out noninvariant states
from the theory with local gauge symmetry of such a type.
Thus, because of the identification (\ref{boundcond})
all physical observables should be invariant under transformation
of the local $Z(3)$ group. All noninvariant states, like single quark
states, must be projected out from the theory to obtain a true Hilbert
space. This defines an ensemble canonical with respect to triality.
On the other hand, it is a natural extension of the orbifold construction
for nonabelian gauge theories at finite temperature.

There is no difficulty to demonstrate that the summation over all
boundary conditions (\ref{boundcond})
is identical with the projection onto the zero triality
sector of the fermionic determinant (a similar proof that only quark
loops of zero triality  contribute to the partition function in this
case has also been given in \cite{Polonyi}). We can define the triality
operator $\hat{\cal{T}}$ (with eigenvalues $0, \pm 1$) which determines the
triality of closed loops, and with his help a projection operator
$\hat{P}_{\cal T}$ on triality $\cal T$
\begin{equation}
\hat{P}_{\cal T} = \frac{1}{3} \sum_{k=0,\pm 1} e^{k 2 \pi i (\hat{\cal T}
- {\cal T})/3}.
\end{equation}
The implementation of
this projection operator in the lattice formulation is
very simple. One has to multiply all links in time direction for
an arbitrary time step with the phase $e^{k 2 \pi i /3}$ and to determine
with this new link variables the determinant. The result has to be
multiplied with $e^{k 2 \pi i {\cal T}/3}$. Finally, the sum over  phases
$k = 0, \pm 1$ has to be executed.
Then the triality zero contribution of the fermionic determinant is
\begin{equation}
{\rm det}_0 M = \hat{P}_0 {\rm det} M = \frac{1}{3} \sum_{k=0,\pm 1} e^{k 2
\pi i \hat{\cal T}/3} {\rm det} M
\end{equation}
and the full fermionic determinant is the sum over all triality sectors
\begin{equation} \label{detM}
{\rm det} M = \sum_{{\cal T}=0,\pm 1} {\rm det}_{\cal T} M,
\end{equation}
with
\begin{equation}
{\rm det}_{\cal T} M = \hat{P}_{\cal T} {\rm det} M.
\end{equation}
It can be easily seen that this procedure is exactly equivalent to the
summation over all boundary conditions (\ref{boundcond}).
{}From eq. (\ref{detM}) it is obvious that the observable $L$ having
triality ${\cal T}=1$ tests only the ${\cal T}=-1$ sector of the fermionic
determinant which is not the vacuum sector.
Using the correct ${\cal T}=0$
vacuum sector for the fermionic determinant results in a $Z(3)$ symmetric
$L$-distribution. Let us discuss this in more detail.
As the measure ${\cal D} [U]$ and the gluonic Lagrangian $S_G[U]$
are $Z(3)$ symmetric, the product ${\rm det} M[U] O_{\cal T}[U]$ in
\begin{equation} \label{observable}
<O_{\cal T}> = \frac{1}{Z} \int {\cal D} [U] e^{-S_G[U]} {\rm det} M[U]
O_{\cal T}[U]
\end{equation}
has to be also $Z(3)$ symmetric to get $<O_{\cal T}> \ne 0$. In other words
$Z(3)$ violating contributions in ${\rm det} M[U] O_{\cal T}[U]$ are
automatically eliminated by the path integral. Only the triality $-{\cal
T}$ component ${\rm det}_{-{\cal T}} M$ of the fermionic determinant
survives. ${\rm det}_{-{\cal T}} M$ has ${\cal T}$ quark loops less than
antiquark loops winding around the lattice in time direction.

Applying this statement to the example of the single Polyakov loop $L$
representing an infinitely heavy quark $Q$, the fermionic determinant
supplies a triality $-1$ state of light quarks which will mostly consist of
a light antiquark $\bar{q}$ in order to colour neutralize the heavy quark $Q$.
The Polyakov expectation value $<L>$ is a
thermodynamical mixture
\begin{equation}
<L> = ... + e^{-F(Q\bar{q})/T} + e^{-F(Qqq)/T} + ...
\end{equation}
of a heavy-light meson $Q\bar{q}$, a heavy-light baryon $Qqq$, ... . These
heavy-light states should be  bound in the hadronic phase and ''ionized'' in
the
quark-gluon plasma phase. The charge density of such light quarks around
static sources has been measured in ref. \cite{Mueller}.

For the system of a heavy quark $Q$ at position $\vec{r}_1$ and a heavy
antiquark $\bar{Q}$ at position $\vec{r}_2$
\begin{equation}
L(\vec{r}_1) L^*(\vec{r}_2)
\end{equation}
a totaly different component of the fermionic determinant, the triality
zero component, contributes. It is evident that this component is $Z(3)$
symmetric, i.e. the three $Z(3)$ transformed sectors which differ in $L$ by
factors $e^{\pm \frac{ 2 \pi i}{3}}$ contribute with the same Boltzmann
factor. This example shows that for observables of triality zero full QCD
is $Z(3)$ symmetric. The fermionic action does not destroy $Z(3)$ symmetry
if we choose to work in the present scheme.

\section{$Z(3)$ symmetry in the Hamiltonian formulation}

It may be instructive to see how the $Z(3)$ symmetry enters the
Hamiltonian formulation of QCD. In temporal gauge we have to integrate
over all time-independent gauge transformations of gauge and of matter
fields. This defines a projection operator
\begin{equation}
P = \int D[\alpha_{\vec{x}}] \exp [-i\int d \vec{x}
\alpha^a(\vec{x})((DE)^a - \rho^a) + i\int_s d\vec{s}\alpha^aE^a],
\label{Hpr1}
\end{equation}
\noindent
where $\rho^a=\bar{\Psi}_{\vec{x}} (\lambda^{a}/2) \Psi_{\vec{x}}$.
$P$ projects onto "physical" states which have to fulfil a local
Gauss law: $(DE)^a=\rho^a$ and a global one: $\int_V divE^a dV=Q^a=0$.
The global Gauss law warrants that physical states are global colour
singlets. We would like to
concentrate here on the centre of the global gauge transformations.
In this case the operator (\ref{Hpr1}) takes the form
\begin{equation}
P = \delta ((DE)^a - \rho^a) P_{tr}
\label{Hpr2}
\end{equation}
\noindent
which includes a Kronecker-delta for the quark triality charge
\begin{equation}
P_{tr} = \frac{1}{N} \sum_{k=0}^{k=N-1}
\exp [\frac{2 \pi ik}{N} (\sum_{x} \bar{\Psi}_{x} \lambda^{a} \Psi_{x})]
\label{q0tr}
\end{equation}
\noindent
(where $\lambda^{a} = \sqrt{3} \lambda^{8}$ for $SU(3)$) and, thus, defines
canonical ensemble with respect to triality. Following the usual
procedure \cite{gross} one can return to the Lagrangian formulation
with a summation over all boundary conditions (\ref{boundcond})
for fermionic fields.

Instead of the canonical ensemble with $B={\cal T}=0$ which should be well
suited, a grand canonical description is allowed if it predicts the same
behaviour in the thermodynamical limit. Common lattice Monte-Carlo
calculations in full QCD use the full fermionic determinant ${\rm det} M$
and thus simulate a grand canonical ensemble with chemical potential
$\mu=0$ containing all  three triality sectors.
We want to point out that there are neither requirements nor
conditions in the theory of strong interactions with long-range gauge fields
guaranteeing that these two descriptions should coincide \cite{Hagedorn,DeTar}.
One of the main conclusion of ref. \cite{Hagedorn} is that one must verify
the agreement between ensembles in each particular case (i.e., for each
conserved quantum number) separately.  It has been proved \cite{Hagedorn}
for free theories that canonical and grand canonical ensembles are
identical in the thermodynamical limit with respect to the baryonic number
and the strangeness. A similar proof for the global colour charge
has been done in ref.\cite{Gorenstain}.
We have shown here that for the full theory
grand canonical and canonical ensembles differ
with respect to quark triality: the $Z(3)$ symmetry has different
realizations in these two ensembles. In ref.\cite{polonyi1} it was
demonstrated that some of thermodynamical functions may have different
behaviour as well.

The most often objection against the above picture is that the projection
onto the zero triality sector defines a new theory which can somehow differ
from the genuine QCD. We believe that the present consideration avoids this
misunderstanding, which has nothing to do with the discussed situation.
We have shown that this projection is equivalent to using the
canonical ensemble with respect to triality in the Hamiltonian formulation.
Further, the Lagrangian formulation of the theory showed that we did not
modify the QCD action.
The canonical ensemble description is realized in this formulation by
means of a summation over the corresponding boundary conditions.

\section{Phase structure of full QCD}

The present consideration allows us to reexamine a possible phase
structure of full QCD and to speculate about the high temperature region
within the canonical ensemble description.
One may claim by now that a quantum field system with a nonabelian local
gauge symmetry including discrete (local and/or global) $Z(N)$ symmetry
can be in one of the following states depending on the concrete
representations of gauge and matter fields in the initial Lagrangian:

1. Coulomb phase, gauge fields are massless.

2. Higgs phase of the first type, triality screening phase.
Gauge symmetry is completely broken, gauge fields acquire masses.

3. Higgs phase of the second type, all colour charges excepting
nonzero triality are screened. Gauge symmetry is broken up to its
discrete centre, no massless gauge fields. Triality charges are
screened classically but may be detected at long distances by specific
quantum mechanical process.

4. Confinement phase. Triality charges are unscreened in any sense but
cannot be detected at long distances. Gauge symmetry is broken either
up to its $Z(N)$ subgroup (vortex mechanism of confinement) or up to
its abelian subgroup (monopole mechanism).

We would like to conjecture the following scenario to be realized in QCD
with dynamical quarks in the fundamental representation:

1. QCD at zero temperature always is in the confining phase.
It seems this statement is generally accepted by now although
there is no strict proof of this fact.

2. Full QCD at finite temperature, as described in the
grand canonical ensemble, always is in the Higgs phase of the first type
where also triality charges are screened.
The aperiodic $Z(3)$ symmetry is explicitly
broken at any temperature and domains with different phases
contribute in different ways to the partition function.
In the thermodynamical limit only one domain with fixed phase
($k=0$ in the commonly used description) can be found.

3. Full QCD at finite temperature, as described in the
canonical ensemble, is in the confining phase at low temperature.
When the temperature increases a phase transition
to the Higgs phase of the second type takes place.
To decide what phase is realized in both ensembles one can apply
an order parameter proposed in \cite{ds3}. This
order parameter can probe a phase of a matter field which carries
$Z(N)$ charge. In the GCE only one domain can be found in the
thermodynamical limit. It implies that the corresponding order parameter
will be always equal to $1$ ($k=0$ phase). In the CE all domains with
different phases are degenerate (see discussion just below)
and we should expect that this order parameter can distinguish between
different phases in the thermodynamical limit. This means that triality
charges of quarks can be detected at long distances via the Aharonov-Bohm
effect \cite{ds1,ds3}.

One of the interesting consequences of the canonical description concerns
the possible states of quark-gluon matter at high temperature.
Recently, in refs.  \cite{Dixit_Ogilivie,Kajantie} it has been concluded on
the one hand that explicit $Z(3)$ symmetry breaking leads to the existence
of metastable states at arbitrary high temperatures.  On the other hand the
authors of refs. \cite{Weiss,Chen} believe that $Z(3)$ phases cannot be
prepared as real macroscopic systems \cite{Gocksch}. These models are based
on the $Z(3)$ asymmetric $L$-distribution in the quark-grand canonical
ensemble. But as we argued above one has to investigate the zero-triality
sector of the fermionic determinant in order to study the QCD vacuum. In
the corresponding ensemble one obtains $Z(3)$ symmetric $L$-distributions
and hence no metastable phases can be found which persist up to infinitely
high temperatures.  The effective models of refs.
\cite{Dixit_Ogilivie,Kajantie,Weiss,Chen} are effective models of
heavy-light hadrons in the QCD vacuum and not effective models of ''the''
QCD vacuum. Exact two-loop calculations in CE, which demonstrate that
all $Z(N)$ phases are left degenerate in CE, have recently been presented
in \cite{ol}. It is just this degeneracy which gives the possibility
to observe different $Z(N)$ domains via the Aharonov-Bohm effect.

\vglue 0.6cm
\noindent
Authors wishes
to thank J.~Engels, C.~DeTar, S.~Olejnik, M.~Oleszczuk, B.~Petersson,
J.~Polonyi, K.~Redlich and H.~Satz for interesting discussions.
The work was partially supported by BMWF and NATO Linkage Grant No.930224.


\begin{thebibliography}{99}
\baselineskip=14pt
\bibitem{preprint} M.~Faber, O.~Borisenko, S.~Mashkevich, G.~Zinovjev,
Fresh look on triality, IK-TUW-Preprint 9209401 (1992) and
9308401 (1993);  \\
Triality in QCD, Proc. of Int. School-Seminar'93
"Hadrons and Nuclei from QCD" (ed.by K.Fyjii, Y.Akaishi, B.L.Reznik.
World Scientific Singapore-New-Jersey-London-Hong-Kong) p.333.
\bibitem{L94} Triality and the grand canonical ensemble in QCD, M.~Faber,
O.~Borisenko, S.~Mashkevich, G.~Zinovjev, Proc. of "Lattice'94", Nucl. Phys. B
(Proc. Suppl.), in print.
\bibitem{Polonyi}  M.~Oleszczuk and J.~Polonyi, Canonical vs. grand
canonical ensemble in QCD, TPR 92-34, 1992.
\bibitem{polonyi1}  M.~Oleszczuk and J.~Polonyi, Ann. Phys. 227 (1993)
76.
\bibitem{smilga} A.V.~Smilga, Ann. Phys. 234 (1994) 1.
\bibitem{kogan} Ian I.~Kogan, Phys. Rev. D 49 (1994) 6799.
\bibitem{hansson} T.H.~Hansson, H.B.~Nielsen and I.~Zahed, QED with unequal
charges, a study of spontaneous $Z_{n}$ symmetry breaking, Preprint
USITP-94-09, SUNY-NTG-94-23, 1994.
\bibitem{kiskis} J.~Kiskis, Phase of the Wilson line, Preprint UCD 94-21,
1994.
\bibitem{boor} J.~Boorstein, D.~Kutasov, Wilson loops, winding modes
and domain walls in finite temperature QCD, Preprint EFI-94-42,
HEP-TH-9409128.
\bibitem{Mac_Lerran} L.~D.~McLerran and B.~Svetitsky, Phys. Rev. D 24
(1981) 450; B.~Svetitsky and L.G.~Yaffe, Nucl. Phys. B 210 (1982) 423;
J.~Kuti, J.~Polonyi, K.~Szlachanyi, Phys. Lett. 98 B (1981) 199
\bibitem{ds1} L.M.~Krauss and F.~Wilczek, Phys. Rev. Lett. 62 (1989) 1221;
T.~Banks, Nucl. Phys. B 323 (1989) 90.
\bibitem{ds2} K.~Li, Nucl. Phys. B 361 (1991) 437.
\bibitem{ds3} J.~Preskill, L.M.~Krauss, Nucl. Phys. B 341 (1990) 50.
\bibitem{ds4} S.~Coleman, J.~Preskill, F.~Wilczek, Nucl. Phys. B 378
(1992) 175.
\bibitem{staggered_fermions} J.~B.~Kogut, Rev. Mod. Phys. 55 (1983) 755.
\bibitem{Banks} T.~Banks and A.~Ukawa, Nucl. Phys. B 225 (1983) 145.
\bibitem{mack} G.~Mack, Phys. Lett. B 78 (1978) 263.
\bibitem{Multiplets} P.~H.~Damgaard, Phys. Lett. B 194 (1987) 107;
H.~Markum, M.~Faber, M.~Meinhart, Phys. Rev. D 36 (1987) 632.
\bibitem{Mueller} W.~B\"urger, M.~Faber, H.~Markum and M.~M\"uller, Phys.
Rev. D 47 (1993) 3034.
\bibitem{gross} D.~Gross, L.G.~Yaffe, R.D.~Pisarski, Rev. Mod. Phys. 53 (1981)
2305.
\bibitem{Hagedorn} R.~Hagedorn, K.~Redlich, Z. Phys. C 27 (1985) 541 (and
references therein).
\bibitem{DeTar} C.~DeTar and L.~D.~Mc Lerran, Phys. Lett. B 119 (1982) 171.
\bibitem{Gorenstain} M.~Gorenstein, S.~Lipskikh, V.~Petrov, G.~Zinovjev,
Phys.Lett. B 123 (1983) 437; H.-Th ~Elze, W.~Greiner, J.~Rafelski,
Phys.Lett. B 124 (1983) 515.
\bibitem{Dixit_Ogilivie} V.~Dixit and M.~C.~Ogilvie, Phys. Lett. B 269
(1991) 353.
\bibitem{Kajantie} J.~Ignatius, K.~Kajantie and K.~Rummukainen, Phys. Rev.
Lett. 68 (1992) 737.
\bibitem{Weiss} V.~M.~Belyaev, Ian~I.~Kogan, G.~W.~Semenoff and N.~Weiss,
Phys. Lett. B 277 (1992) 331.
\bibitem{Chen} W.~Chen, M.~I.~Dobroliubov and G.~B.~Semenoff, Phys. Rev. D 46
(1992) R1223.
\bibitem{Gocksch} For a discussion concerning this point see also
A.~Gocksch and R.~D.~Pisarski, Nucl. Phys. B 402 (1993) 657.
\bibitem{ol} M.~Oleszczuk, Degenerate $Z(2)$ domains in the $SU(2)$
quark canonical ensemble, TPR-94-11, 1994.

\end{thebibliography}
\end{document}